# Flexible Sensory Platform Based on an Electrolyte-Gated Oxide Neuron Transistor

Ning Liu[1,2], Li Qiang Zhu[2], Ping Feng[1], Chang Jin Wan[2], Yang Hui Liu[2], Yi Shi[1] & Qing Wan[1]

Abstract: Inspired by the dendritic integration and spiking operation of a biological neuron, flexible oxide-based neuron transistors gated by solid-state electrolyte films are fabricated on flexible plastic substrates for biochemical sensing applications. When a quasi-static dual-gate laterally synergic sensing mode is adopted, the neuron transistor sensor shows a high pH sensitivity of ~105 mV/pH, which is higher than the Nernst limit. Our results demonstrate that single-spike dynamic mode can remarkably improve the pH sensitivity, reduce response/recover time and power consumption. We also find that appropriate depression applied on the sensing gate electrode can further enhance the pH sensitivity and reduce the power consumption. Our flexible neuron transistors provide a new-concept sensory platform for biochemical detection with high sensitivity, rapid response and ultralow power consumption.

[1]School of Electronic Science & Engineering, Nanjing University, Nanjing 210093, People's Republic of China.

[2]Ningbo Institute of Materials Technology and Engineering, Chinese Academy of Sciences, Ningbo 315201, People's Republic of China.

Correspondance and requests for materials should be addressed to Q. W or Y. S (email: wanqing@nju.edu.cn; yshi@nju.edu.cn)

With the recent interest in brain/computer interfaces[1], soft robotics[2], wearable electronics[3] and skin-like sensory systems[4], flexible devices have attracted widespread attention. These emerging devices require new fabrication schemes that enable integration with soft, curvilinear and time-dynamic human tissues. Among these devices, flexible sensors are becoming increasingly significant in a wide-variety of novel applications such as in vivo monitoring[5], delivery of advanced therapies[6], artificial sense organs[7], etc. As a fundamental component for sensor application, field-effect transistors (FETs) based sensors have been intensively investigated due to their inherent advantages of miniaturization, facilitated integration, direct transduction and label-free detection[8-11]. The classical sensing mechanism of the FET-based sensor is attributed to a charge-dependent interfacial potential due to the adsorption of potential-determining species at sensing membrane/electrolyte interface[12]. The sensitivity is limited to ~59.2 mV/decade (Nernst limit) at room temperature when the threshold voltage ($V_{th}$) is recorded as the output signal. It should be noted here that the above mentioned measurements are based on the quasi-static electrostatic coupling mode, which potentially increases the time consumption and energy dissipation. But in smart sensory platforms, such as implantable devices and wearable sensory systems, low power consumption is one of the most important pre-requisites.

Synergic integration of presynaptic inputs from the dendrites plays an important role for sensory information process and cognitive computation, and the idea of building bio-inspired solid-state devices has been around for decades[13,14]. In 1992, Shibata et al proposed the Si-based neuron transistors with multiple input gates that are capacitively coupled to a floating gate[15]. The "on" or "off" state of the neuron transistors depends on the integrated effect of the multiple input gates. One of the unique features of the neuron transistors is the ultralow power dissipation during the calculation due to the gate-level sum operation in a voltage mode. From then on, Si-based neuron transistors have attracted much attention for chemical and biological detection due to the easy adjustment of threshold voltage [16-20]. But, up to now, flexible electrolyte-gated neuron transistors with amorphous oxide channel layers for biochemical sensing applications are not reported.

Amorphous oxide-based transistors were proposed as promising fundamental unit in sensory platform due to their low process temperature, superior electrical properties, high reliability and easy reproducibility[21-23]. To date, remarkable sensing performances have been demonstrated in these oxide-based transistors [24-26]. For portable applications, low-voltage operation is preferred. Electrolyte gated electric-double-layer (EDL) transistors can act as potential candidates with a low operation voltage due to the strong EDL modulation at the electrolyte/channel interface[27,28]. Recently, oxide-based EDL transistors gated by solid-state inorganic electrolytes were proposed by our group[29,30]. At the same time, artificial synapses and neuron transistors with low power consumption and fundamental biological functions were mimicked in individual device[31-33]. In this communication, flexible sensory platform based on individual protonic/electronic coupled indium-zinc-oxide (IZO) neuron transistor was fabricated on plastic substrates. Such neuron transistor exhibited a high sensitivity beyond Nernst limit when a quasi-static dual-gate synergic modulation mode was adopted. Most importantly, single-spike dynamic sensing of the neuron transistor was also investigated, and pH sensing with ultra-high sensitivity, very quick response/recover time, and extremely low power consumption were realized.

Figure 1a shows the schematic diagram of a flexible IZO-based neuron transistor with multiple in-plane gate electrodes for pH sensing application. A miniature Ag/AgCl reference electrode immersed into a 5.0 μL pH buffer solution droplet on the nanogranular $SiO_2$ (n-$SiO_2$) electrolyte film is acted as the sensing gate ($G_1$). In-plane aluminum (Al) electrodes ($G_2$, $G_3$…$G_n$) are used as the control gates. The distinctive feature of our neuron transistor is that sensing gate and all control gates in the same plane can be electrostatic capacitively coupled to the bottom floating gate. The capacitive network of the neuron transistor is plotted in Fig. 1b. The carrier density of the IZO channel layer can be electrostatic modulated by the weighted sum of all inputs of the sensing and control gates. The weight for each gate is directly proportional to the capacitive factor normalized by the total capacitance of the floating gate[5]. Figure 1c displays the top-view optical image of the IZO-based neuron transistor sensor in practice. The channel width (W) and length (L) is 1000 and 80 μm, respectively. As a proof of concept, only one control gate electrode is used in this work. The distance between the control gate electrode and the drain electrode is 300 μm. Figure 1d shows a picture of the IZO-based neuron transistor array on the PET plastic substrate, exhibiting its flexible nature under external force.

Figure 2a shows the transfer characteristics of the IZO neuron transistor at a constant $V_{DS}$ of 1.5 V. The gate voltage applied on lateral Al gate electrode is swept from −1.5V to 1.5V and then back. A clear anticlockwise hysteresis window of ~0.4 V is observed, which is likely due to the mobile protons in the n-SiO$_2$ electrolyte[34]. The subthreshold swings (SS), current on/off ratio ($I_{on}/I_{off}$) and $V_{th}$ are estimated to be ~175 mV/decade, ~6.4×10$^5$, and −0.3 V, respectively. In addition, the field-effect electron mobility ($\mu_{FE}$) at the saturation region is estimated to be ~12 cm$^2$/V·s by the following equation:

$$I_{DS} = \frac{WC_i\mu_{FE}}{2L}(V_{GS}-V_{th})^2 \quad (1)$$

where $C_i$ (~2.7 μF/cm$^2$) is the specific capacitance of the n-SiO$_2$ electrolyte film measured from two in-plane Al gate electrodes at 1.0 Hz (Supplementary Fig. S1). For practical flexible electronics application, flexible devices should be bendable without sacrificing their electrical properties. The influence of mechanical bending on the electrical characteristics of our neuron transistors was investigated. Figure 2b shows the transfer ($I_{DS}$–$V_{G2}$) curves recorded before, during and after bending by a cylinder with a radius of 1.0 cm. The images of the measurement process are shown in the insets of Fig. 2b. Good reproducibility is obtained on different test conditions. Moreover, mechanical stress tests have also been performed by bending the sample repeatedly. Figure 2c shows the transfer curves recorded at repetitive bending cycles. The flexible neuron transistors survive after more than 1000 flex/flat cycles with negligible change in the transfer characteristics. The variations in $V_{th}$ and $\mu_{FE}$ with the repetitive bending cycles are extracted in Fig. 2d. After 1000 cycles of bending and recovery, a small positive shift of ~0.1 V in $V_{th}$ and only 10% reduction in $\mu_{FE}$ are measured. The results indicate that the flexible neuron transistors have good mechanical reproducibility and durability.

We will next study the pH sensing performance of the devices operated in the quasi-static mode. Figure 3a shows the transfer curves of the neuron transistor sensor operated in the linear region ($V_{DS}$=0.1 V) when the sensing gate is immersed into solution droplets with different pH values. The inset in Fig. 3a shows the layout of this normal pH sensing measurement. Clear negative shift of the transfer curve is observed when pH value is changed from 10 to 4. It has been reported that acidic solution can give rise to a more positive surface potential due to the ionic interaction at the solution/$SiO_2$ interface[35,36]. In our case, positive surface potential will make protons in n-$SiO_2$ electrolyte migrate to the electrolyte/IZO channel interface, which will induce excess electrons in the IZO channel and a negative shift of transfer curve. When the gate voltage at a drain current of 10 nA is defined as the responsive voltages ($V_R$), a sensitivity of ~37.4 mV/pH is realized, as shown in Fig. 3b. This value is comparable to the reported FET sensors using $SiO_2$ as a sensing material[37].

In order to improve the sensing performance of the IZO-based neuron transistor, dual-gate gate synergic modulation mode is investigated, where $G_1$ is biased at different fixed voltages and $G_2$ is swept from −2.0 to 1.0 V. The schematic of the measurement is shown as Fig. 4a. Figure 4b shows the transfer curves ($I_{DS}$-$V_{G2}$) curves measured at $V_{DS}$=0.1 V when pH value is changed from 10 to 4, where $V_{G1}$ is fixed at 0.3 V and −0.6 V, respectively. Similarly, the transfer curves shift to the negative direction when the pH value is decreased at a fixed $V_{G1}$. Here we should point out that more obvious shifts in the transfer curve are induced by pH variation when $V_{G1}=-0.6$ V. The sensitivity in terms of $V_R$ shift is plotted as a function of $V_{G1}$ (Fig. 4d). The pH sensitivity increases when $V_{G1}$ shifts from a positive value to a negative value. A maximal pH sensitivity of ~105 mV/pH is obtained when $V_{G1}=-0.6$ V. This value is much larger than the Nernst limit of ~59.2 mV/pH. The improved sensitivity obtained at a negative $V_{G1}$ is attributed to asymmetric and synergic capacitive coupling between these two gates ($G_1$ and $G_2$). Figure 4c shows real-time responses of $I_{DS}$ of the IZO-based neuron transistor sensor in different pH solutions (pH=10, 8, 6, 4) for 180s at fixed $V_{DS}$=0.1V, $V_{G1}$=0 V and $V_{G2}$=0.2V. It is observed that $I_{DS}$ increases gradually to a stable value in an equilibrium time of ~20 s at each pH value due to the effect of protonic migration. The steady $I_{DS}$ increases stepwise with discrete change in pH value from 10 to 4. The sensitivity S of a sensor can also be defined as the relative change in channel conductance, $S = (|G-G_0|)/(G_0)=\Delta G/G_0$ [38]. In our case, the response conductance to pH=10 is defined as $G_0$. Therefore, the sensitivity ($\Delta G/G_0$) is estimated to be ~2.2 for pH=4 at equilibrium state. We also find that the sensitivity ($\Delta G/G_0$) can be improved by a negative bias applied on sensing gate ($G_1$), as shown in the right axis of Fig. 4d. A highest sensitivity of ~38.3 is obtained at a fixed $V_{G1}$ of –0.6 V. This value is much higher than those reported in nanoscale transistor sensors[39,40]. This is because an appropriate negative voltage applied on the sensing gate (G1) can make the neuron transistor operated in the subthreshold regime, which is favorable for sensitivity enhancement, as reported in nanowire FET biosensors [41]. At last, we should point out that synergic modulation is not limited to two gate structure. Neuron transistors with multiple gates can be easily fabricated for our in-plane gate structure.

Next, inspired by the spiking operation mode of a biological neuron, we will investigate the single-spike pH sensing performance of our neuron transistor sensors. Due to the distinctive dynamic characteristics of the proton migration, our device presents a unique time dependent transient property. During the measurement, equilibrium is disturbed by a small voltage pulse applied on the control gate ($G_2$). The dynamic spike current response to such a disturbation contains the pH sensing information. After the detection, the device will quickly recover to the original equilibrium state. Moreover, during the single-spike sensing process, the energy consumption is extremely low, which is preferred for portable and wearable sensory applications. The single-spike pH sensing measurement of the detection is schematically illustrated in Fig. 5a. At first, a disturbing spike $V_{G2}$ (0.2 V, 10 ms) was applied on control gate ($G_2$), and a synchronous reading spike $V_D$ (0.02 V, 10 ms) was applied on drain electrode to measure the output current. As shown in Fig. 5b, when the pH value is changed from 4 to 10, the response current ($I_{DS}$) decreases from 512 to 80 nA. We also find that the logarithm of $I_{DS}$ decreases linearly with increasing pH value, and a high sensitivity ($\Delta G/G_0$) of ~5.6 is estimated, as shown in Fig. 5c. The response/recover time is estimated to be several milliseconds, which is much shorter than that operated in quasi-static mode. The reproducibility of the single-spike pH sensing measurement is also investigated. Figure 5d shows the response currents stimulated by repeated voltage pulse spikes in the case of pH=6. The results indicate a good reproducibility of single-spike detection of pH values. The energy dissipation of our system can be estimated by multiplying the reading voltage, the channel current and the spike duration time [42]. Figure 5e shows the average energy dissipation for single-spike pH detection in each pH value from pH=10 to pH=4 with a spike duration time of 10 ms.. The energy dissipation reduces from 103 pJ/spike to 15.6 pJ/spike when the pH value is increased from 4 to 10. Of course, the energy dissipation can be reduced further by decreasing spike voltage and reducing spike duration time. The influence of bending on the sensitivity is also investigated. As shown in Fig. 5f, after 1000 bending cycles, the sensitivity reduction is less than 10% for both quasi-static and single-spike sensing modes.

The single-spike sensing performance implemented with an asynchronous reading spike is also investigated. Figure 6a shows the sensitivity as a function of the inter-spike interval ($\Delta t$) between $V_D$ and $V_{G2}$. If $\Delta t<0$, the reading spike $V_D$ is applied before $V_{G2}$, the protonic disturbance does not happen in the sensing process, thus the sensitivities are close to the equilibrium state and a sensitivity ($\Delta G/G_0$) of ~2.7 is obtained. When $\Delta t \geq 0$, the measured sensitivity is time interval ($\Delta t$) dependent. A highest pH sensitivity ($\Delta G/G_0$) of 5.6 is be measured when $\Delta t=0$, and it gradually reduces to 2.7 with increasing $\Delta t$. Figure 6b shows the sensitivity as a function of the spike time duration. We find out that the sensitivity decreases gradually to ~2.4 when the spike duration is increased to 2000 ms. These results indicate that the neuron transistor sensor tends to arrive at equilibrium state with the increase of spike duration. Hence, higher sensitivity can be obtained if the spike time duration is further reduced (Supplementary Note 1). The sensitivity of single-spike pH sensing performance of our neuron transistor sensor can be further enhanced by another gate synergic modulation. Figure 6 c shows the influence of voltage bias applied on $G_1$ on the sensitivity when the device is operated in single-spike mode. The pH sensitivity increases when $V_{G1}$ shifts from positive to negative. A maximum sensitivity ($\Delta G/G_0$) of ~63 can be obtained when a negative voltage of –0.2 V is applied on $G_1$. We also investigated the influence of voltage bias applied on $V_{G1}$ on the energy dissipation of single-spike sensing measurement. Our results indicate that the energy dissipation can be gradually reduced when the $V_{G1}$ is changed from 0.2V to -0.2V. As shown in Fig. 6 d, an ultra-low energy dissipation of 0.6 pJ/spike is estimated for pH=10 at $V_{G1}$= -0.2 V when the spike time duration is 10 ms. Similar to the quasi-static synergic mode, a appropriate negative $V_{G1}$ can make the device operate in the subthreshold regime, and then a enhanced sensitivity can be obtained. At the same time, negative bias can reduce the spike sensing current, which is critical for energy dissipation reduction.

In summary, flexible oxide-based neuron transistor sensors were fabricated on plastic substrates. A pH sensitivity of ~105 mV/pH was obtained for quasi-static dual-gate synergic sensing mode. Our results demonstrated that single-spike dynamic operation mode could remarkably improve the pH sensitivity, reduce response/recover time and power consumption. We also found that appropriate depression applied on the sensing gate could further enhance the pH sensitivity and reduce the power consumption. Our results provided a novel strategy for fabricating biochemical sensors with high sensitivity, rapid response and ultralow power consumption.

**Methods**

**Fabrication of flexible oxide-based neuron transistor.** First, 500-nm-thick nanogranular $SiO_2$ (n-$SiO_2$) electrolyte films were deposited on the ITO-coated PET substrates by plasma enhanced chemical vapor deposition (PECVD) at room temperature. $SiH_4$ (95% $SiH_4$+5% $PH_3$) and pure $O_2$ were used as reactive gases. Then, 30-nm-thick IZO channel layer was deposited on the n-$SiO_2$ electrolyte films by magnetron sputtering using a nickel shadow mask. The sputtering was performed at a RF power of 100 W and a working pressure of 0.5 Pa using an IZO target. The channel width and length were 1000 and 80 μm, respectively. Finally, 100-nm-thick Al source/drain electrodes and in-plane gate electrodes were deposited by thermal evaporation patterned by another shadow mask.

**Preparation of pH solution.** pH solutions were prepared by titrating 10 mM phosphate solution with dilute hydrochloric acid or potassiumhydroxide solutions. The pH value of the solutions was monitored by a commercial pH meter. All chemicals were purchased from Sinopharm Chemical Reagent Co., Ltd. (China).

**Electrical and sensing characterizations.** The sensing area of the device was immersed in deionized water for 24 hours before the measurement. Frequency-dependent capacitances of the n-$SiO_2$ electrolyte films were characterized by a Solartron 1260A Impedance Analyzer in air ambient with a relative humidity of ~55%. Transistor characteristics and pH sensing performance were recorded by a semiconductor parameter characterization system (Keithley 4200 SCS) at room temperature. After each pH value test, the solution droplet was removed and the sensing area was rinsed two times in deionized water.

**Acknowledgements**

This work was supported in part by the National Science Foundation for Distinguished Young Scholars of China (Grant No. 61425020), and in part by a Project Funded by the Priority Academic Program Development of Jiangsu Higher Education Institutions, and in part by the Zhejiang Provincial Natural Science Fund (LR13F040001).

**Author contributions**

The manuscript was prepared by N. L., L. Q. Z., F. P. and Q. W. Device fabrication was fabricated by N. L. and Y. H. L. Measurements were performed by N. L. and C. J. W. The project was guided by Q. W. and Y. S.

**Additional information**

**Supplementary Information** accompanies this paper on

http://www.nature.com/naturecommunications

**Competing financial interests**: The authors declare no competing financial interests.

**Figure Legends**

**Figure 1 | IZO-based neuron transistor on PET substrate and its flexibility exhibition.** (a) Schematic of the flexible pH sensor based on an IZO neuron transistor with multiple gate electrodes. An Ag/AgCl reference electrode immersed in the solution droplet is the sensing gate. In-plane Al electrodes are used as the control gates. (b) The schematic image of the capacitive network of the flexible IZO-based neuron transistor. The carrier density of the IZO channel is modulated by the weighted sum of all inputs of sensing gate and control gates. (c) An optical microscope image of the system. (d) The sensor array fabricated on a flexible PET substrate.

**Figure 2 | Electrical properties the IZO-based neuron transistor and its flexibility characteristics.** (a) Transfer curves of the IZO-based neuron transistor measured by sweeping the voltage on the control gate ($G_2$) at $V_{DS}$=1.5 V. An anticlockwise hysteresis loop of ~0.4 V is observed. (b) Transfer curves of the flexible IZO-based neuron transistor measured before, during and after bending by a cylinder with a radius of 1.0 cm. The inset is the pictures during the measurement process. (c) Transfer curves of device measured before and after repeated bending cycles by sweeping the control gate ($G_2$) at $V_{DS}$ = 0.1 V. (d) The variations in $V_{th}$ and $\mu_{FE}$ of the flexible IZO-based neuron transistor with repetitive bending cycles. Error bars represent standard deviations for 5 samples.

**Figure 3 | The pH sensitivity of the IZO-based neuron transistor measured in single $G_1$ sweeping mode.** (a) Transfer curves of the IZO-based neuron transistor measured by using the sensing gate $G_1$ at $V_{DS}$ = 0.1 V. The pH value of the solution droplet on the sensing gate $G_1$ is changed from 4 to 10 at a step of 2. The inset shows the measurement schematic. (b) The sensitivity in terms of $V_R$ shift. The data can be fitted linearly by the black line. A sensitivity of 37.5 mV/pH and a linearity of 0.995 are obtained.

**Figure 4 | The pH sensitivity of IZO-based neuron transistor sensor measured in**

**dual-gate synergic modulation mode. (a)** The schematic image of the measurement. **(b)** Transfer curves of the device measured by applying sweep voltage on the control gate ($G_2$) at $V_{DS}$=0.1 V with different fixed voltages applied on $G_1$. The voltage applied on $G_1$ is 0.3 and −0.6 V. the pH value of the solution droplet on $G_1$ is changed from 4 to 10 at a step of 2. **(c)** The real-time responses of $I_{DS}$ for IZO neuron TFT sensors in each pH solution (pH=10, 8, 6, 4) for 180s at constant $V_{DS}$=0.1V, $V_{G1}$=0V and $V_{G2}$=0.2V. **(d)** The sensitivity in terms of $V_R$ shift and $\Delta G/G_0$ at different $V_{G1}$.

**Figure 5 | The pH sensing performance of IZO-based neuron transistor operated in a single-spike mode. (a)** Schematic of single-spike pH sensing measurement. **(b)** Single-spike measurement is performed as the pH value of the solution changed from 4 to 10. The spike voltage $V_{G2}$ (0.02V, 10 ms) and the reading voltage $V_D$ (0.2 V, 10 ms) are synchronous voltage pluses. The reference electrode $V_{G1}$ is grounded. **(c)** The logarithm of $I_{DS}$ peak changes linearly with the pH value of the solution. The error bars represent standard deviations for 10 samples. **(d)** Reproducibility of the neuron transistor sensor for pH=6 solution. **(e)** pH value dependent energy dissipation operated in single-spike mode. **(f)** The influence of 1000 times bending on the sensitivity of the neuron transistor for both quasi-static and single-spike sensing modes. Fixed biases ($V_{DS}$=0.1 V, $V_{G1}$=0 V, $V_{G2}$=0.2 V) are applied in quasi-static mode. Synchronous pulse voltages $V_{G2}$ (0.2 V, 10 ms) and $V_D$ (0.02 V, 10 ms) with fixed $V_{G1}$ bias of 0 V are applied in dynamic spiking mode.

**Figure 6 | The influence of measuring parameters on the spike sensing performance. (a)** The sensitivity as a function of the inter-spike interval between $V_D$ and $V_{G2}$ for asynchronous spiking sensing test. **(b)** The changes in sensitivity with spike duration. The solid line is a fitted curve. **(c)** The changes in sensitivity and energy dissipation (pH=10) with various $V_{G2}$ spike amplitude for synchronous spiking sensing test. **(d)** The changes in sensitivity and energy dissipation (pH=10) against various $V_{G1}$ bias.

**Figures**

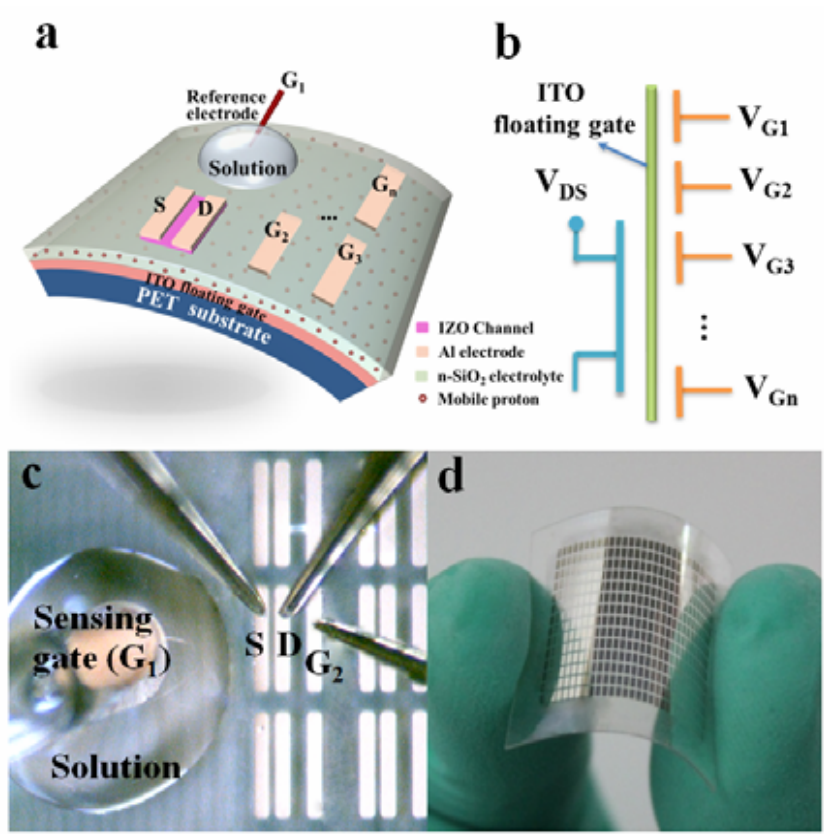

Figure 1

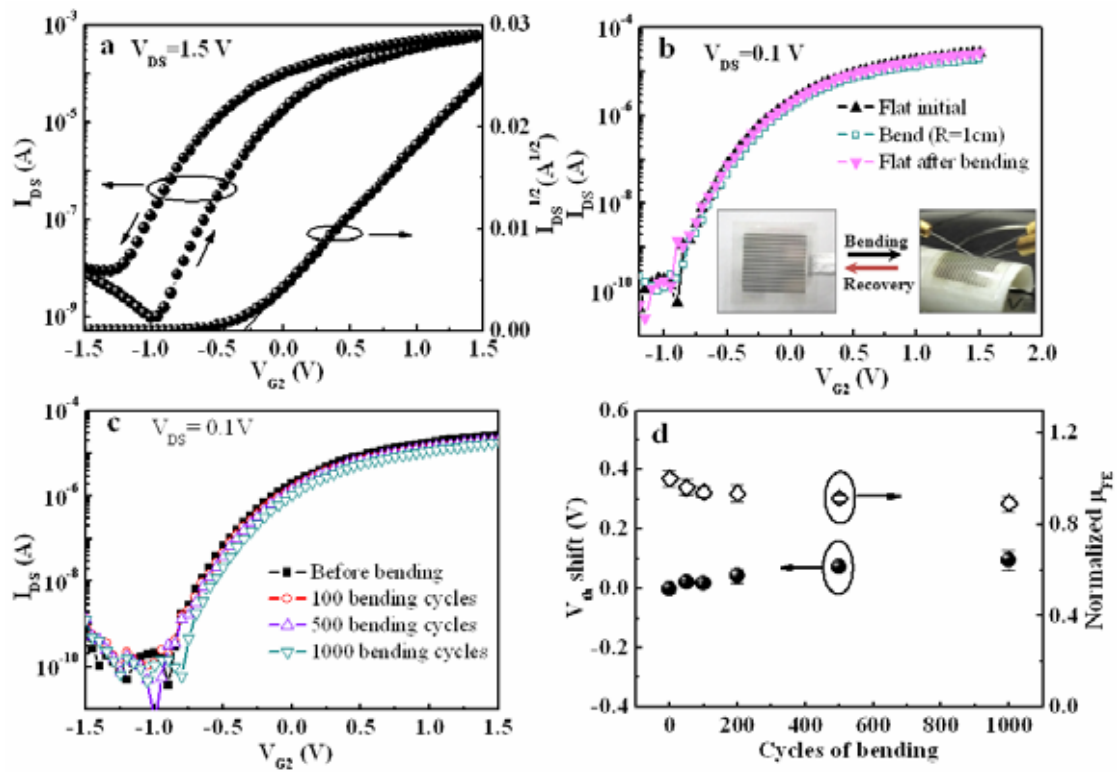

**Figure 2**

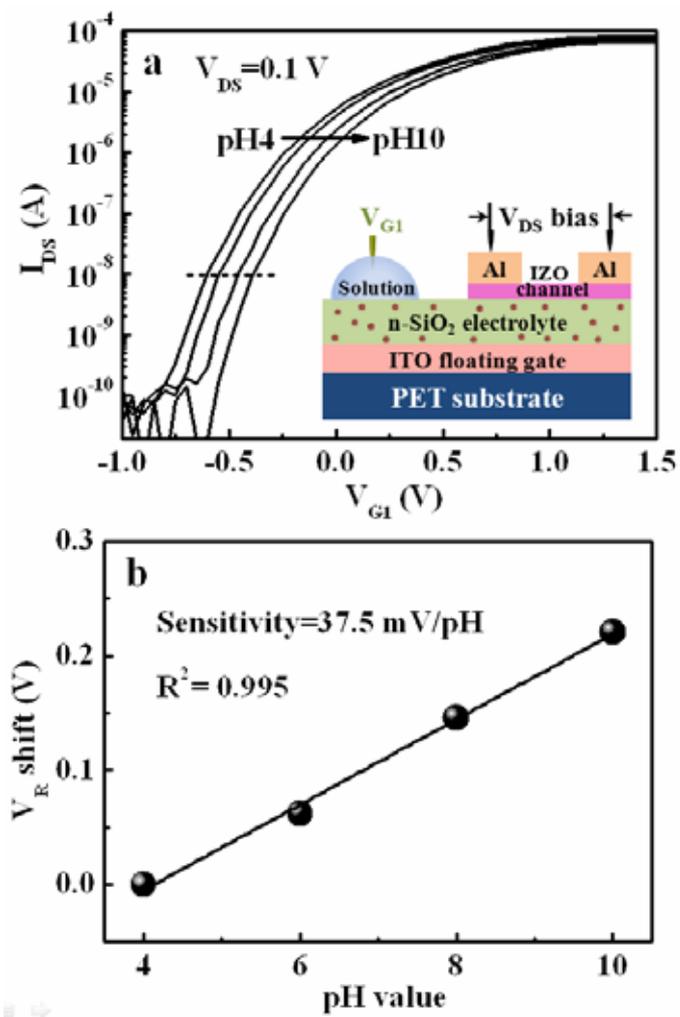

Figure 3

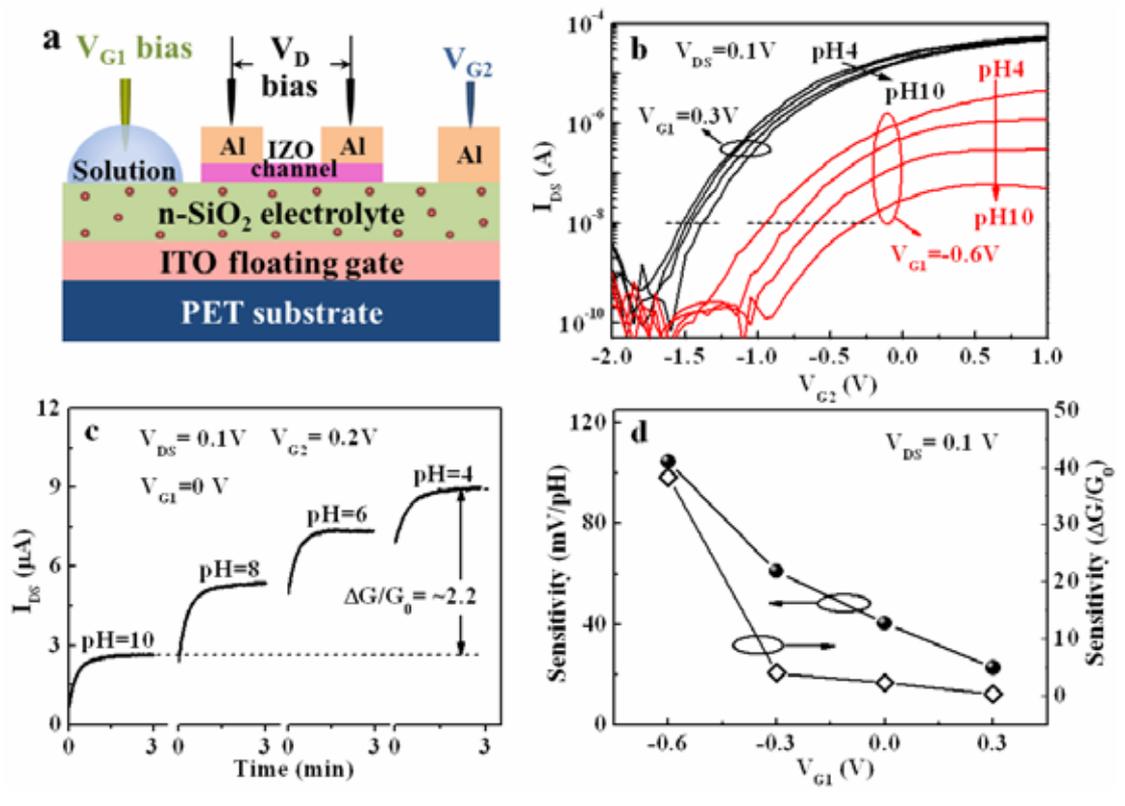

**Figure 4**

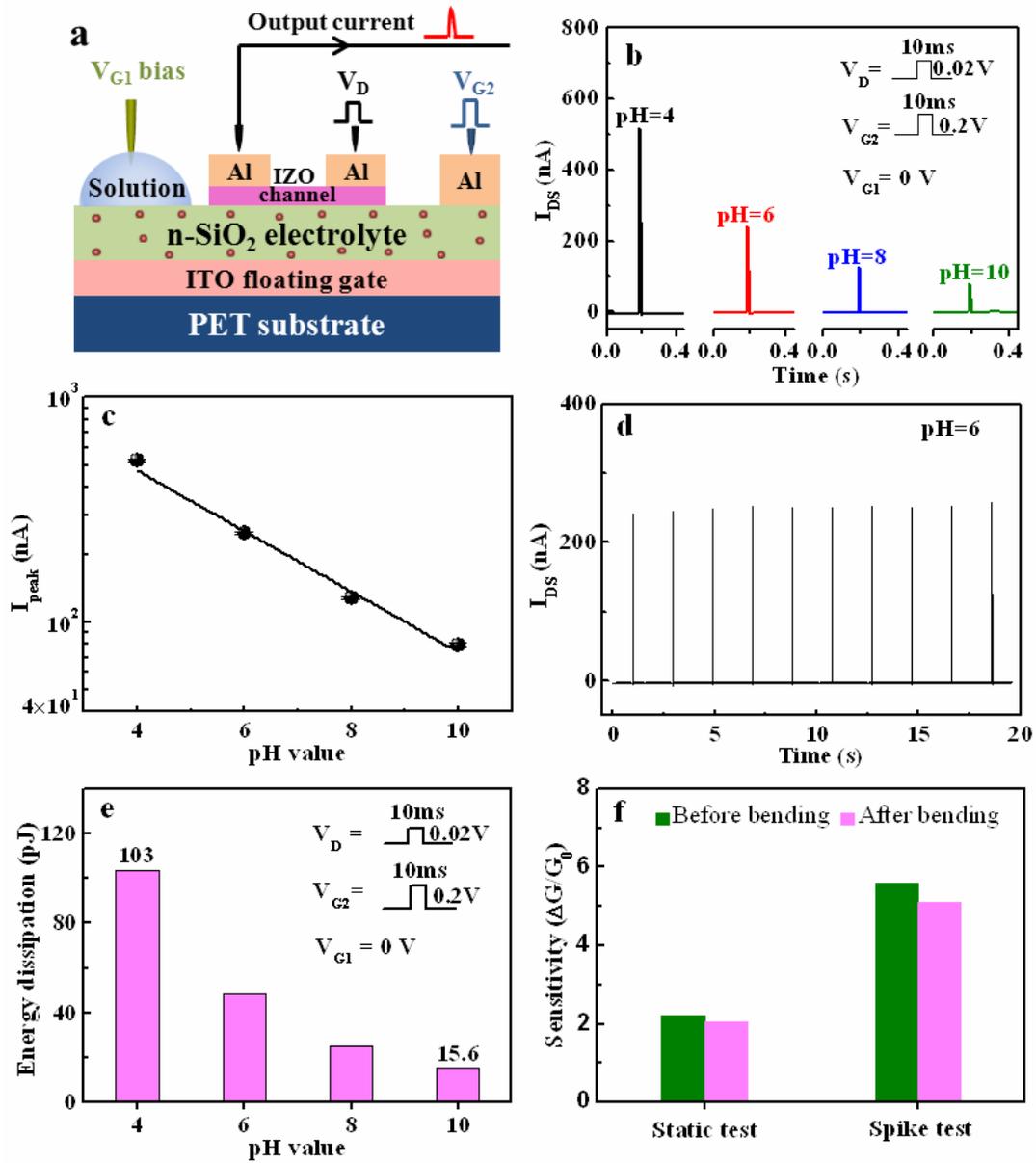

**Figure 5**

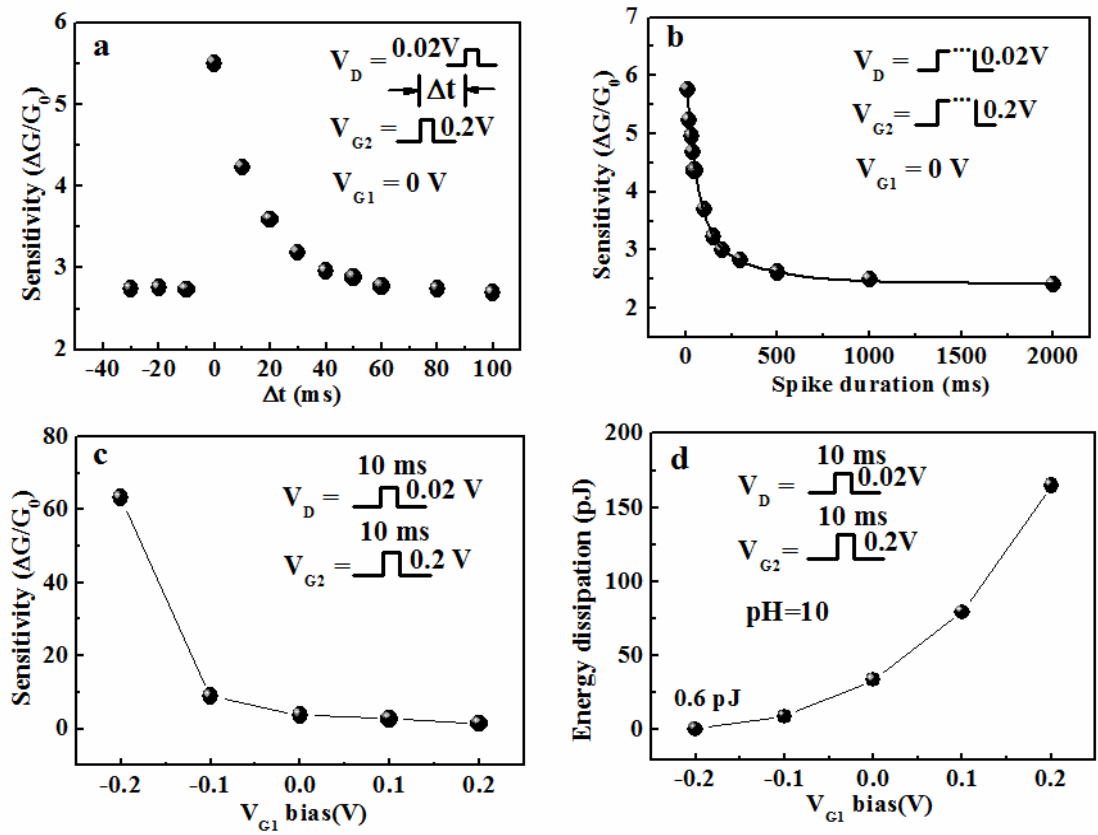

**Figure 6**

**Supplementary Information**

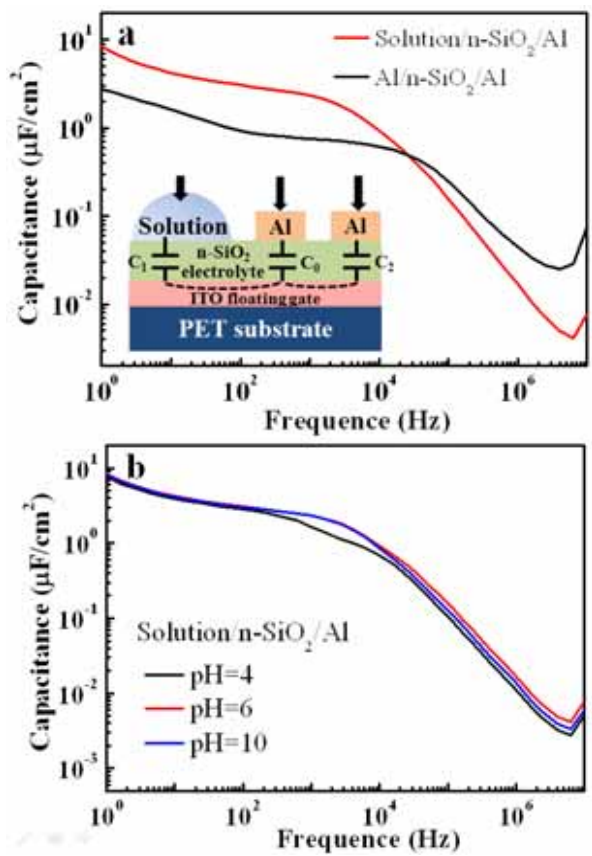

**Supplementary Figure S1 | Capacitance-frequency curves of the n-SiO$_2$ electrolyte films.** **(a)** The specific capacitance–frequency curves for the n-SiO$_2$ electrolyte films on PET substrate with a liquid droplet (pH=6) and an Al electrode, using an in-plane test structure. Inset shows the schematic diagram of the capacitance measurement. The specific capacitances at 1.0 Hz are estimated to be ~8.3 μF/cm$^2$ and ~2.7 μF/cm$^2$ for liquid droplet and Al electrode, respectively. **(b)** The specific capacitance–frequency curves for the n-SiO$_2$ electrolyte films gated by sensing gate with different pH solutions (pH= 4, 6, 10). Little difference between the C-f curves at different pH solutions is observed.

**Supplementary Note 1. Theoretical analysis of single-spike pH sensing process**

According to the celebrated site-binding theory and Gouy-Chapman-Stern model[1,2], the $V_{th}$ shift of FET-based sensor results from a change in surface potential ($\varphi_0$) at the interface between the solution and ion sensing membrane. The relation between the surface potential ($\varphi_0$) and the pH value of the solution is given by:

$$\frac{d\varphi_0}{dpH} = -2.3\alpha\frac{kT}{q} \tag{1}$$

Where $k$ is the Boltzmann constant, T is the absolute temperature, $q$ is the elementary charge, and $\alpha$ is a dimensionless sensitivity parameter which varies between 0 and 1. When the sensitivity parameter α approaches to unity, the sensing membrane shows a maximum pH sensitivity of 59.2 mV/pH at room temperature (298 K), which is known as the Nernst limit.

For neuron transistor sensors, asymmetric capacitive coupling between the control gate and sensing gate to floating gate results in intrinsic amplification of the measured surface potential shifts [3,4]. In our dual-gate operation mode, the $V_{th}$ shift of neuron transistor sensor can be deduced as[5,6]:

$$\Delta V_{th} = -\frac{C_1}{C_2}\Delta\varphi_0 = -\eta \cdot \Delta\varphi_0 \tag{2}$$

where $C_1$ and $C_2$ is the specific capacitance of sensing gate and control gate, respectively. $\eta$ is the factor representing the coupling ratio between these two gates.

For Field-effect transistors, the current is exponential to the gate voltage in the subthreshold region,

$$I_{DS} \sim \exp[\frac{q(V_{GS} - V_{th})}{kT}] \tag{3}$$

Referring to equation (2) and (3), the sensitivity of dual-gate transistor sensors in terms of $\Delta G/G_0$ at equilibrium state can be deduced as:

$$\frac{\Delta G}{G_0} = \frac{G_a - G_b}{G_b} = \exp(\frac{\eta q \Delta \varphi_0}{kT}) - 1 \tag{4}$$

Where $G_a$ and $G_b$ are the channel conductance of the neuron transistor sensors measured with acidic and alkaline solution, respectively. $\Delta \varphi_0$ is the change in surface potential induced by pH variations. $\eta$ is the factor representing the coupling ratio between the sensing gate and control gate. In the spike sensing mode, the channel current is defined by the ionic diffusion related to the EDL coupling effect. The EDL capacitor can be modeled as an ideal resistor-capacitor (RC) circuit. The effective voltage on the EDL capacitor is given by:

$$V_E = V_{GS}[1 - \exp(-\frac{t}{\tau})] \tag{5}$$

where t is the voltage bias duration, $\tau$ is the time constant of equivalent RC circuit of the EDL capacitor. According to (3) and (5), the subthreshold current is

$$I_{DS} \sim \exp\{\frac{qV_{GS}(1-\exp(-t/\tau)) - qV_{th}]}{kT}\} \tag{6}$$

Thus, the sensitivity in terms of $\Delta G/G_0$ for the spiking mode is

$$\frac{\Delta G}{G_0} = \exp(\eta q \Delta \varphi_0/kT) \cdot \exp\{\frac{qV_{GS}}{kT}[\exp(-t/\tau_b) - \exp(-t/\tau_a)]\} - 1 \tag{7}$$

where $\tau_a$ and $\tau_b$ are the time constant of the EDL capacitor for the neuron transistor sensor under acidic and alkaline solution, respectively. Defining a proportion function $f(t,\tau) = \exp(\frac{qV_{GS}}{kT}[\exp(-t/\tau_b) - \exp(-t/\tau_a)])$, the sensitivity in terms of $\Delta G/G_0$ for spiking mode can be approximated to be:

$$\frac{\Delta G}{G_0} = f(t,\tau) \cdot \exp(\eta q \Delta \varphi_0/kT) - 1 \tag{8}$$

For a short spike duration, if $\tau_a < \tau_b$, $f(t, \tau) > 1$, the sensitivity is enhanced. If $\tau_a > \tau_b$, $f(t, \tau) < 1$, the sensitivity is reduced. For a long spike duration, $f(t, \tau) \rightarrow 1$, it is equal to the sensing process at equilibrium state.

The value of $\tau$ reflects the transporting speed of protons in n-SiO$_2$ electrolyte films. At a lower pH value, acidic solution gives rise to a more positive surface potential, which promotes the transporting of protons within n-SiO$_2$ electrolyte films, and leads to a smaller $\tau$ value. Hence, an amplified factor ($f(t, \tau) > 1$) for pH sensitivity can be realized in the spike sensing mode. The $\tau$ values can be deduced by fitting the transient current of IZO neuron transistor sensors as a function of time, as shown in the Fig. S2.

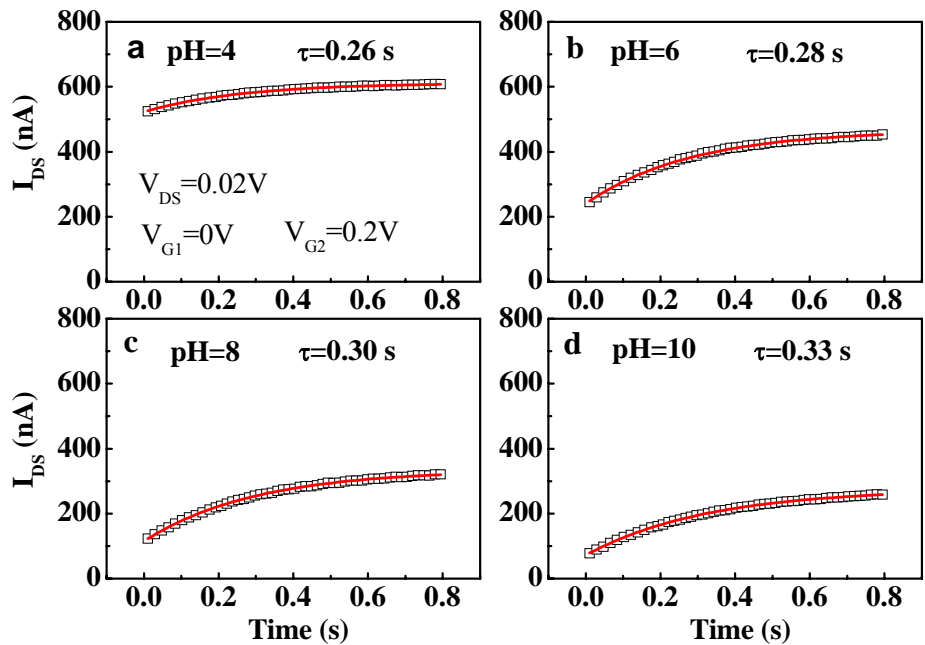

**Supplementary Figure S2 | The transient current versus time for different pH values.** (a) pH=4; (b) pH=6; (c) pH=8; (d) pH=10. Fixed voltage biases ($V_{DS}$=0.2V, $V_{G1}$=0V, $V_{G2}$=0.2V) are applied. The transient currents can be well fitted by equation (6), as shown in the red solid lines. The fitted $\tau$ is estimated to be 0.26s, 0.28s, 0.30s and 0.33s for pH value of 4, 6, 8 and 10, respectively.